
\magnification=1200
\hoffset=0.1truein
\voffset=0.1truein
\vsize=23.0truecm
\hsize=16.25truecm
\parskip=0.2truecm
\def\newpage{\vfill\eject}

\def\pp{\parshape 2 0.0truecm 16.25truecm 2truecm 14.25truecm}

\def\ypro{ {\mu} }

\def\scale{ {\rm A} }
\def\pot{ {\psi_{\rm ex}} }

\def\dpdrho{ {\partial p \over \partial \rho}  }
\def\gtwid{\mathrel{\raise.3ex\hbox{$>$\kern-.75em\lower1ex\hbox{$\sim$}}}}
\def\ltwid{\mathrel{\raise.3ex\hbox{$<$\kern-.75em\lower1ex\hbox{$\sim$}}}}
\vskip 0.5truein
\centerline{\bf GENERAL ANALYTIC RESULTS FOR NONLINEAR WAVES}
\centerline{\bf  AND SOLITONS IN MOLECULAR CLOUDS}
\vskip 0.2truein
\centerline{\bf Fred C. Adams, Marco Fatuzzo$^\dagger$, and Richard Watkins}
\vskip 0.1truein
\centerline{\it Physics Department, University of Michigan}
\centerline{\it Ann Arbor, MI 48109}
\vskip 0.4truein

\centerline{\it submitted to The Astrophysical Journal: 25 May 1993}
\centerline{\it revised: 17 September, 1993}

\vskip 0.4truein

\bigskip
\centerline{\bf ABSTRACT}
\medskip

We study nonlinear wave phenomena in self-gravitating
fluid systems, with a particular emphasis on applications
to molecular clouds.  This paper presents analytical results
for one spatial dimension. We show that a large class of
physical systems can be described by theories with a
``charge density'' $q(\rho)$; this quantity replaces the
density on the right hand side of the Poisson equation for
the gravitational potential.  We use this formulation to
prove general results about nonlinear wave motions in
self-gravitating systems.  We show that in order for
stationary waves to exist, the total charge (the integral
of the charge density over the wave profile) must vanish.
This ``no-charge'' property for solitary waves is related
to the capability of a system to be stable
to gravitational perturbations for
arbitrarily long wavelengths.  We find necessary and
sufficient conditions on the charge density for the
existence of solitary waves and stationary waves.
We study nonlinear wave motions for Jeans-type theories [where
$q(\rho) = \rho - \rho_0$] and find that nonlinear waves
of large amplitude are confined to a rather narrow range
of wavelengths.  We also study wave motions for molecular
clouds threaded by magnetic fields and show how the allowed
range of wavelengths is affected by the field strength.
Since the gravitational force in one spatial dimension does
not fall off with distance, we consider two classes of models
with more realistic gravity: Yukawa potentials and a
pseudo-two-dimensional treatment.  We study the allowed
types of wave behavior for these models.   Finally, we
discuss the implications of this work for molecular
cloud structure.  We argue that molecular clouds can
support a wide variety of wave motions and suggest that
stationary waves (such as those considered in this paper)
may have already been observed.

\vskip 0.4truein
\noindent
{\it Subject headings:} hydromagnetics -- wave motions --
interstellar: molecules -- stars: formation

\vskip 0.4truein
${}^\dagger$ Compton GRO Fellow

\newpage
\bigskip
\centerline{\bf 1. INTRODUCTION}
\medskip

Molecular clouds are self-gravitating fluid systems and
are capable of supporting a wide variety of oscillatory
motions.  The overall goal of this work is to understand the
dynamics of volume density waves in molecular clouds and to
explore the extent to which waves can explain the observed
cloud structure. These clouds comprise a substantial fraction of
the gas in the galaxy, they provide the initial conditions for
star formation, and they are interesting astrophysical
objects in their own right.  Molecular clouds are now fairly
well-observed and exhibit very complicated highly nonlinear
structures (see, e.g., the recent review of Blitz 1993).
Thus far, however, no definitive theory exists to explain or
predict the observed large scale structure of these clouds.
Our working hypothesis is that volume density wave motions can
produce (at least in part) this complicated structure.

The study of nonlinear waves and solitons began over a century ago
(see the classic papers by Russell 1844 and Riemann 1858).  Similarly,
the study of gravitational instability in astrophysical fluids has
had a long and distinguished history (see Jeans 1928).  In the
meantime, however, remarkably little work has been done on the
intersection of these two subjects, i.e., the study of nonlinear
waves and solitons in systems where self-gravity is important.
Liang (1979) has studied nonlinear waves in the cosmological fluid
and has shown that soliton solutions are not allowed for ordinary
pressure laws (see also Ray 1983).  G\"otz (1988) has searched
for solitons in Newtonian gravity, although the solutions found
there are not relevant for our present discussion.  Other related
previous work does not directly address the problem of nonlinear
wave motions in a self-gravitating fluid.  Nonlinear Alfv\'en waves
have been studied and have been proposed as a mechanism for producing
clumpy structure in clouds (Elmegreen 1990), although this study does
not include self-gravity.  Much of the previous work has been devoted
to the linear stability of clouds (see, e.g., Langer 1978;
Pudritz 1990; Dewar 1970; see also the review of Shu, Adams, \& Lizano
1987).  On a smaller size scale, the formation of molecular cloud cores
has been considered through the process of ambipolar diffusion
(Mouschovias 1976, 1978; Shu 1983; Nakano 1985; Lizano \& Shu 1989;
Shu et al. 1987).  On the size scale of molecular clouds, the
study of wave motions including self-gravity remains
largely untouched.

Self-gravity is an important ingredient for the existence of
stationary nonlinear waves in a neutral fluid.
In the absence of self-gravity, acoustic waves in fluids are known
to steepen and shock (see, e.g., Whitham 1974; Shu 1992).  On the
other hand, self-gravity provides dispersion (Jeans 1928) which
tends to spread out wave packets.  Thus, fluids with self-gravity can,
in principle, reach a balance between nonlinear steepening and
gravitational dispersion.  This balance leads to the possibility
of nonlinear stationary waves and solitary waves,
which we study in this paper.

In an earlier paper (Adams \& Fatuzzo 1993; hereafter Paper I),
we began a study of nonlinear waves and solitons in molecular
clouds.  We found one class of nonlinear waves for molecular
clouds with no magnetic fields; we also showed that no soliton
solutions exist for these systems.  The one dimensional system
of Paper I (with no magnetic fields) is highly idealized and
hence unphysical in the following ways:
(1) The usual Poisson equation for gravity in one dimension
produces a gravitational force that does not fall off at large
distances, i.e., there is no asymptotic regime where the
gravitational force goes to zero.
(2) The one dimensional formulation of the problem does not
allow for any two (or three) dimensional effects.
(3) Magnetic fields are ignored, but are known to be dynamically
important.  (In Paper I, we briefly considered clouds with magnetic
fields and derived a model equation which allows a wide variety of
solutions, including nonlinear waves, solitons, and topological
solitons -- see \S 2.3 below).
(4) Real (observed) molecular clouds are not collapsing as
a whole, whereas the one dimensional model does collapse.
In other words, the one dimensional model does not allow
for a static equilibrium state.
We note that items (1) and (2) are coupled in the sense that any
physical system will have a finite spatial extent in the direction
perpendicular to the wave motion.  This finite extent implies a
finite mass and hence allows gravity to fall off with distance.
Similarly, items (3) and (4) are also coupled. Magnetic fields
are (at least in part) responsible for the fact that clouds are
not collapsing as a whole.

This present paper is a generalization of Paper I.
Since the full magnetohydrodynamic problem is rather complicated
and the chances of finding meaningful analytic solutions seem
remote,  we use the full physical problem as motivation to
derive a collection of model equations which approximate the
true behavior of the system.
In order to obtain analytic results, some of these
approximations are by necessity rather severe.  However, our
goal is to obtain a qualitative physical understanding of the
behavior of nonlinear waves in self-gravitating fluids.
This approach has the advantage of allowing
for analytic results which in turn provide us with a clear
picture of ``what depends on what''.  Since we must sacrifice
quantitative accuracy in our model equations, this work should
be complemented by both numerical studies and by studies in higher
spatial dimensions.  We are currently pursuing these avenues of
research.

This paper is organized as follows.  We begin with a general
formulation of the problem in \S 2.  We develop the concept
of a ``charge density'' where we replace the right hand side
of Poisson's equation for the gravitational field with a
more complicated function (this concept was introduced in
Adams, Fatuzzo, \& Watkins 1993; hereafter AFW).
We show that many types of physical systems
can be modeled in this manner. In \S 3 we prove
general results which hold for all theories with a charge
density.  In particular, we prove that stationary waves must
have zero total charge when integrated over one wavelength.
We also derive necessary and sufficient conditions on the
charge density for the existence of stationary waves and
solitary waves.
In \S 4 we consider the special case where the charge density
$q = \rho - \rho_0$; this case corresponds to the traditional
Jeans analysis. In \S 5, we include the effects of a magnetic
field which is perpendicular to the wave propagation
direction. In \S 6, we study nonlinear waves for
theories where gravity is modeled with a Yukawa potential.
This formulation of the problem allows the gravitational force
to fall off at large distances.
In \S 7, we show how stationary waves in two
spatial dimensions can produce an effective charge density
theory in one spatial dimension.  We conclude in \S 8 with a
discussion and summary of our results.

\bigskip
\centerline{\bf 2. GENERAL FORMULATION FOR ONE-DIMENSIONAL WAVES}
\medskip

In this section, we introduce a class of model equations for the
study of nonlinear waves in molecular clouds.  We begin with
the equations of motion for a molecular cloud fluid.
In one spatial dimension, these equations take the form:
$${\partial \rho \over \partial t} +
{\partial \over \partial x} (\rho u) = 0 ,  \eqno(2.1)$$
$${\partial u \over \partial t} + u
{\partial u \over \partial x} + {1 \over \rho}
{\partial p \over \partial x} +
{\partial \psi \over \partial x} = 0 , \eqno(2.2)$$
$${\partial^2 \psi \over \partial x^2} = 4 \pi G \rho .
\eqno(2.3)$$
For most of this paper, we take the pressure of the molecular
cloud fluid to have a general barotropic form,
$$p = p(\rho)  , \eqno(2.4{\rm a})$$
which includes most equations of state of interest.
In order to obtain explicit results, we sometimes must specify
the form of the pressure law.  In these cases, we adopt the form
$$p = a_s^2 \rho + p_0 \log (\rho/\rho_R)  , \eqno(2.4{\rm b})$$
where the first term corresponds to an isothermal equation of
state ($a_s$ is the isothermal sound speed).  The second term
arises from a ``turbulent'' contribution.  This term is motivated
by the observational finding that linewidths in molecular clouds
vary with density according to $\Delta v \sim \rho^{-1/2}$
(see, e.g., Larson 1981; Myers 1983; Dame et al. 1986; Myers
1987; and Myers \& Fuller 1992).  If we interpret the observed
linewidth as the effective transport speed in the fluid,
we obtain $(\Delta v)^2$ = $v^2_{\rm turb}$ =
$\partial p_{\rm turb} / \partial \rho$ = $p_0 / \rho$,
where the final equality follows from the observed relation.
We then obtain the form given in equation [2.4b] by integration
(see also Lizano \& Shu 1989; Myers \& Fuller 1992).

We find it convenient to work in dimensionless units.
We let $\rho_R$ denote a reference density and $a_s$ denote
the sound speed.
We can non-dimensionalize all quantities according to
$$u \to u/a_s , \eqno(2.5{\rm a})$$
$$\rho \to \rho/\rho_R , \eqno(2.5{\rm b})$$
$$x \to k x \qquad {\rm where}
\qquad k^2 = 4 \pi G \rho_R / a^2_s , \eqno(2.5{\rm c})$$
$$t \to k a_s t , \eqno(2.5{\rm d})$$
$$p \to {p \over \rho_R a_s^2} \, , \eqno(2.5{\rm e})$$
$$\psi\to\psi /a_s^2\, , \eqno(2.5{\rm f})$$
$$\kappa \equiv {p_0 \over a^2_s \rho_R } \, . \eqno(2.5{\rm g})$$
For applications to molecular clouds, we are primarily interested
in spatial size scales of 1 -- 30 pc where number densities are
typically $\sim$100 -- 1000 cm$^{-3}$.  The temperature is expected
to be in the range $T$ = 10 -- 35 K and hence the sound speed
$a_s$ = 0.20 -- 0.35 km/s. The wavenumber $k$ = $2 \pi / \lambda_J$
where $\lambda_J$ is the (usual) Jean's length that would result
from thermal pressure (with sound speed $a_s$ and fiducial
density $\rho_R$).  For example, if we take $a_s$ = 0.2 km/s
and $\rho_R = 2 \times 10^{-22}$ g cm$^{-3}$ (the conditions
roughly appropriate for the Taurus molecular cloud),
we get $x=1$ at a physical length scale of $\sim$0.5 pc.
The parameter $\kappa$ determines the relative size of the
``turbulent''  contribution to the pressure; the aforementioned
observations indicate that $\kappa$ lies in the range 6 -- 50.

\bigskip
\centerline{\it 2.1 The Concept of Charge Density}
\medskip

In this section we generalize the theory described in equations
[2.1 -- 2.4] by introducing the concept of a ``charge density''
(see AFW).  We can include a variety of additional behavior simply
by modifying the Poisson equation to take the form
$${\partial^2 \psi \over \partial x^2} = q ( \rho ) \, ,
\eqno(2.6)$$
where $q(\rho)$ is a function of the density $\rho$.
We denote the quantity $q(\rho)$ as the {\it charge density}.
We use this terminology because, in general, whatever appears
on the right hand side of a Poisson equation is often called
``the charge density''.  Notice that $q(\rho)$ defined
here has absolutely nothing to do with the electric charge density
(which of course is completely negligible in this problem).
Notice also that the choice $q = \rho$ gives us back the usual system.
In the following subsections (\S 2.2 -- 2.4), we present physically
motivated examples which can be written as charge density theories.

One general motivation for introducing a charge density is to
consider modifications of one-dimensional gravity (see below)
and/or to include additional long range forces in the problem.
We can model such systems by assuming the forces are conservative
($\sim$ $\nabla\pot$) and thus can be written in terms of
a new potential $\pot$.
We then consider the potential $\psi$ to be the total
potential, i.e., the sum of the usual gravitational potential
and the new potential $\pot$.  The dimensionless Poisson
equation must have the form
$${\partial^2 \psi \over \partial x^2} =
\rho + {\it something} \equiv q \, , $$
where we have defined the right hand side of the equation to be
the charge density.  Here we consider those special cases
where the charge density is a function of $\rho$ only.

\bigskip
\centerline{\it 2.2 Jeans Theory}
\medskip

The simplest possible nontrivial extension of the theory arises
from subtracting out the contribution to the gravitational potential
due to the background fluid (see, e.g., Jeans 1928;
Binney \& Tremaine 1987).  This approximation can be written
in the form of a charge density theory with
$$q(\rho) = \rho  - \rho_0 \, , \eqno(2.7)$$
where $\rho_0$ is the dimensionless background density of the fluid.
Thus, the traditional Jeans analysis results in a charge density
theory.

One physically motivated way to obtain a theory with the same
mathematical form as in equation [2.7] is to posit a physical system
which rotates at a uniform rate $\Omega$ (see, e.g., Shu 1992).  In
the direction perpendicular to the rotation axis, Gauss's law for
a uniform density cylinder with density $\rho_0$ implies
$$- \varpi \Omega^2 = - 2 \pi G \rho_0 \varpi \, ,  \eqno(2.8)$$
where $\varpi$ is the radial coordinate (note that all quantities in
equation [2.8] have their usual dimensions).   Thus, the uniform
density state will be in mechanical balance provided that
$$\rho_0 = \Omega^2/2 \pi G \, . \eqno(2.9)$$
It is straightforward to show (a homework problem in Shu 1992)
that for waves propagating along the axis of rotation in this
system, we obtain equations of motion with a ``charge density''
of the form $q = \rho - \rho_0$, where $\rho_0$ is given by
equation [2.9].

We now consider ``typical'' values for the rotationally induced
fiducial density $\rho_0$.
At the solar circle, the rotation rate around the center of the
galaxy is of order $\Omega \sim 10^{-15}$ rad/s.  With this value
of $\Omega$, the density implied by equation [2.9] corresponds to
a number density of $n_0 \sim 1$ cm$^{-3}$. In molecular clouds,
the observed rotation rates are larger than this fiducial value by
factors of 10 -- 100 (see, e.g., Goldsmith \& Arquilla 1985) and
hence the implied number density becomes $n_0 \sim 10^2 - 10^4$
cm$^{-3}$. These values thus lie in an interesting range for
molecular clouds.

\bigskip
\bigskip
\centerline{\it 2.3 Charge Density from Magnetic Field Effects}
\medskip

Another way to obtain a charge density is to simulate the effects
of an embedded magnetic field.  In Paper I, we derived a model
equation on this basis. In this model,
we assumed that the magnetic field points in a direction
perpendicular to that of the wave motion and that the
neutral component of the fluid is coupled to the magnetic
field through its frictional interaction with the ionized
fluid component (which is well-coupled to the field).
We also ignored any chemical effects so that the
ionized component of the fluid obeys a continuity
equation. The resulting model equation can be derived from a
theory with the charge density written in the form
$$q (\rho) = \rho + \Gamma \Bigl( {1 \over \rho}
- {1 \over \rho_F} \Bigl) \, , \eqno(2.10)$$
where $\Gamma$ represents the coupling strength between the
neutral and ionized components and where $\rho_F$ determines
the ion density (assumed to be constant in Paper I).  We note
that the model equation corresponding to the charge density
[2.10] is idealized in two ways. First, a dissipative term has
been dropped to obtain this form. Second, the ion density
is not calculated self-consistently.  The net result of these
approximations is to make the magnetic force on the neutral
component into a long-range force that balances the long-range
force of gravity.  This treatment is unphysical in the sense
that these magnetic forces, which arise from the frictional
force exerted on the neutral components by the ions, are
intrinsically local.  On the other hand, this approximation
allows for something to cancel the long-range force of gravity.

\bigskip
\centerline{\it 2.4 Charge Density from Yukawa Potentials}
\medskip

We can also derive a charge density for theories in which
gravity is modeled by a Yukawa potential. The motivation
for this approximation is to include the effects
of a decreasing gravitational field strength while retaining the
one-dimensional treatment of the problem.  As we show here,
Yukawa potentials provide one means of realizing this
behavior. In particular, we write the Poisson equation in
the generalized form
$${\partial^2 \psi \over \partial x^2} =  m^2 \psi + \rho \, .
\eqno(2.11)$$
The Green's function for the operator $\partial^2 / \partial x^2 - m^2$
has an exponential fall off and hence produces an exponential fall off
in the gravitational force between point masses. The value of the
parameter $m$ determines the effective range of the force.
\footnote{$^\dagger$}{For completeness, we note that a more
complicated additional term of the form $m^2 \psi^n$ could be
used; this term represents nonlinear interactions in
the gravitational field and is not of interest here.  Notice,
however, that this nonlinear theory can also be written as a
charge density theory.}

The right hand side of equation [2.11] defines an effective charge
density $q$, although it is not written explicitly
as a function of density only.   We note, however, that we can
integrate the stationary version of the force equation [2.2]
to obtain
$$\psi + h(\rho) + {1\over2} v^2 = E \, , \eqno(2.12)$$
where $E$ is the constant of integration (a discussion of the
stationary wave approximation is given in \S 2.5).  The quantity
$h (\rho)$ is the enthalpy and is defined by
$$h(\rho) = \int^\rho {d p \over \rho} \, . \eqno(2.13)$$
The speed $v$ can be eliminated by using the solution of
the continuity equation for stationary waves,
i.e., $v = A /\rho$ (see equation [2.17] below).
If we now solve equation [2.12] for $\psi$ and use the result in
equation [2.11], we can read off the appropriate form of the
charge density, i.e.,
$$q(\rho) = \rho + m^2 \bigl[ E - h(\rho) - {A^2 \over 2 \rho^2}
\bigr] \, . \eqno(2.14)$$
As we show in \S 7, another way to obtain a one dimensional theory
where gravity falls off with distance is to begin with a two dimensional
theory and then reduce it to a one dimensional theory through the
use of rather severe approximations.  We can thus obtain yet another
charge density theory (compare equations [7.9] and [2.14]).

\bigskip
\bigskip
\goodbreak
\centerline{\it 2.5 Stationary Waves and Solitary Waves}
\medskip

In this paper, we are interested in stationary nonlinear waves
and their limiting forms known as solitary waves and solitons.
Since the definitions of these wave entities vary greatly in the
literature (e.g., Whitham 1974; Coleman 1985; Rajaraman 1987;
Drazin \& Johnson 1989; and Infeld \& Rowlands 1990), we
present the following working definitions from Paper I:
We use the term {\it stationary wave} to mean any
wave structure that is a function of the variable $\xi = x - v_0 t$
only. We use the term {\it solitary wave} to refer to any solution
of a nonlinear wave equation (or system of equations) which
(1) represents a stationary wave, and (2) is localized in space
so that the wave form either decays or approaches a constant at
spatial infinity.  We use the term {\it soliton} to be
synonymous with {\it solitary wave}, although the term {\it soliton}
is often reserved for waves which also satisfy the additional requirement:
(3) the wave form can interact strongly with other solitons and
retain its identity.  Solutions which satisfy requirement (3) must
have extraordinary stability in order to pass through each other
and, after emerging from the collision, retain their initial forms.
Such solitons are very rare; wave entities which satisfy the first
two requirements and not (3) occur much more frequently.

We first consider stationary waves, which correspond to traveling
waves of permanent form. For these waves, the fluid fields are
functions of the quantity
$$\xi = x - v_0 t , \eqno(2.15)$$
where $v_0$ is the (nondimensional) speed of the wave.
Next, we introduce a new velocity variable
$$v = u - v_0 , \eqno(2.16)$$
which is simply the speed of the fluid relative to the speed $v_0$
of the wave. Using the above definitions in the continuity equation
[2.1], we find the relation
$$\rho v = A = {\it constant} , \eqno(2.17)$$
where the constant of integration $A$ is the
``Mach number'' of the wave.

We want to combine the equations of motion to obtain a
single nonlinear differential equation for the density $\rho$.
If we differentiate the force equation [2.2] with respect
to $x$ and use the generalized Poisson equation [2.6]
to eliminate the potential, we obtain
$$\rho \rho_{\xi \xi} \Bigl[ \rho^2 \dpdrho - A^2 \Bigr] +
\rho_\xi \rho_\xi \Bigl[ 3 A^2 - \rho^2 \dpdrho
+ \rho^3 {\partial^2 p \over \partial \rho^2} \Bigr]
+ \rho^4 q (\rho)  = 0 ,  \eqno(2.18)$$
where subscripts denote differentiation and
where we have eliminated the velocity dependence by using
relation [2.17].  The equation of motion [2.18] is the
fundamental equation of this paper.
Notice its highly nonlinear nature.  Fortunately and somewhat
surprisingly, however, we can integrate this equation to obtain
$${1 \over 2} \rho_\xi^2 =  \rho^6
\Bigl[ \rho^2 \dpdrho - A^2 \Bigr]^{-2} f(\rho) \equiv
{\cal F} (\rho) \, , \eqno(2.19)$$
where we have defined $f(\rho)$ to be an integral that depends
on the form of the charge density $q(\rho)$, i.e.,
$$f(\rho)  = \int^\rho  \, d \rho \, {q(\rho)\over \rho} \,
\bigl[ {A^2 \over \rho^2}  - \dpdrho \bigl]  \, .
\eqno(2.20)$$

The existence or non-existence of stationary waves for a particular
physical system can be understood through the methods of phase
plane analysis (see, e.g., Infeld \& Rowlands 1990;
Drazin \& Johnson 1989).  As illustrated above, in this method
we reduce the system of equations to a single equation of the form
$${1 \over 2} \rho_\xi^2 = {\cal F} (\rho, C_j) , \eqno(2.21)$$
where the $C_j$ are constants (see equations [2.19] and [2.20]).
Notice that equation [2.21] is just the virial theorem for an
analogue particle moving in a potential $-{\cal F}$ (see, e.g.,
Rajaraman 1987); when viewed in this manner, equation [2.21] shows
that the properties of the ``potential'' $\cal F$ determine the
allowed behavior of the analogue particle and hence the properties
of the wave solutions $\rho(\xi)$ in a fairly simple manner.
In particular, the form of ${\cal F} (\rho)$ determines whether
or not solitary wave solutions can exist.
We also note that if dissipative terms are present in the original
equation of motion, then the solution cannot be written in the form
of equation [2.21] and hence stationary wave solutions do not exist.

We first note that physically meaningful solutions must have
${\cal F} \ge 0$ (so that the solutions are real).  For the
systems considered here, the field $\rho$ is a mass density and must
always be positive.  Thus, physically relevant solutions exist when
${\cal F}(\rho)$ is positive over a range of postive densities.
Wave solutions exist when ${\cal F}(\rho)$ is positive between
two zeroes of the function $\cal F$, where the zeroes of $\cal F$
correspond to maximum and minimum densities of the wave profile.
The nature of the zeroes of $\cal F$ determines the nature of the
wavelike solutions.  For example, if $\cal F$ is positive between
two simple zeroes, then (ordinary) nonlinear waves result.
However, if $\cal F$ is positive between a simple zero and a
double zero or higher order zero (i.e., any point where both $\cal F$
and $\partial {\cal F}/\partial \rho$ vanish), then a new type of
solution -- a solitary wave -- can arise.  Suppose we expand
equation [2.21] about the double zero, which we denote as $\rho_\infty$;
we obtain
$$\rho_{\xi}^2 = (\rho - \rho_\infty)^2 {\cal F}^{''} (\rho_\infty)
+ {\cal O} \bigl[ (\rho - \rho_\infty)^3 \bigr]  , \eqno(2.22)$$
where ${\cal F}^{''} (\rho_\infty) > 0$ because we are considering
the case in which $\cal F$ is positive.
Thus, as $\rho \to \rho_\infty$, the wave profile has the form
$$\rho - \rho_\infty \sim \delta \exp \Bigl[ \mp
\sqrt{ {\cal F}^{''}(\rho_\infty) }
\xi \Bigr] , \eqno(2.23)$$
where $\delta$ is a constant.  We see that a soliton consists
of a single large hump of material and that the density smoothly
approaches its asymptotic value $\rho_\infty$ as $\xi \to \pm \infty$.
Formally, the wavelength of the solution diverges for a
solitary wave.

Another interesting type of behavior can arise when the
function $\cal F$ is positive between {\it two} double zeroes
of $\cal F$, say $\rho_A$ and $\rho_B$.  In this case, the
solution $\rho (\xi)$ can approach one value (e.g., $\rho_A$)
in the limit $\xi \to -\infty$ and the other value ($\rho_B)$
in the limit $\xi \to \infty$.  Solutions of this type are
known as {\it kinks} or {\it topological solitons} and are
well studied in the context of quantum field theory (e.g.,
Coleman 1985; Rajaraman 1987).

\bigskip
\centerline{\bf 3. GENERAL RESULTS}
\medskip

In this section we present general analytic results that
apply to all theories of the form described in \S 2.
These results are applicable for arbitrary forms of the charge
density $q(\rho)$.  In particular, we state and prove four
elementary ``Theorems'' which greatly constrain the allowed types
of wave behavior for charge density theories.  In the subsequent
sections, we use these results to study the behavior of
theories which contain specific forms for the charge density.

\bigskip
\centerline{\it 3.1 The No-Charge Property for Stationary Waves}
\medskip

To begin, we present an argument which shows that for any
theory with a charge density $q(\rho)$, a strong constraint must
be met in order for physically relevant stationary waves to exist.
This constraint arises from the fact that a stationary wave must
have local extrema where the pressure and velocity gradients go
to zero.  Therefore, in order for the wave to remain stationary,
the force of gravity must also vanish at these points. This
behavior can only occur if the integral of the charge density
$q(\rho)$ between two extrema is zero.  This argument can
be stated as follows:

\proclaim Result 1. If a stationary wave solution $\rho(\xi)$
exists for the one-dimensional theory, then the integral of the
charge density over one wavelength must vanish, i.e.,
$$Q \equiv \int_{-\lambda/2}^{\lambda/2} \, d \xi\, q
\bigr[ \rho (\xi) \bigl] \, = 0 \, , $$
where the wavelength of a soliton is taken to be infinite.
This claim is limited to the case of nonsingular solutions;
for solitons we also require that the density does not vanish
at spatial infinity.
\par

\noindent
In order to show that Result 1 is true, we first
define the total charge $Q$ contained in one wavelength to be
$$Q \equiv \int_{-\lambda/2}^{\lambda/2} \, q \, d\xi =
2 \int_0^{\lambda/2} \, q \, d\xi \, , \eqno(3.1)$$
where $\lambda\to \infty$ for soliton solutions.
The second equality holds because of the symmetry of the
problem about $\xi = 0$ (this result is valid for all classes
of solutions except topological solitons, for which the formula
for $Q$ will be different by a factor of two).
The first integral of the equation of motion for a stationary wave can be
written in the form
$$\rho_\xi  =\pm {\sqrt 2}\rho\left[{\partial p\over\partial\rho}
-{A^2\over\rho^2}\right]^{-1} f^{1/2} \, , \eqno(3.2)$$
where the function $f$ (defined in equation [2.20]) must be positive
definite for valid solutions.
Using the relation
$$d(f^{1/2}) = {1 \over 2} f^{-1/2} {q\over\rho} \left[
{A^2 \over \rho^2} - {\partial p \over \partial \rho} \right]
d\rho \, , \eqno(3.3)$$
along with equation [3.2], we find the identity
$$q \, d\xi  = \pm {\sqrt 2} \, d(f^{1/2}) \; . \eqno(3.4)$$
Equation [3.4] may be substituted into equation [3.1], where
care must be taken to ensure that the integrand has the
proper sign. Using this result, we find
$$Q = 2{\sqrt 2}\int_{\rho_1}^{\rho_2} d (f^{1/2}) \, , \eqno(3.5)$$
which can be integrated to obtain
$$Q = 2{\sqrt 2} \Biggl[f(\rho_2)^{1/2}-f(\rho_1)^{1/2}
\Biggr] = 0 \, .  \eqno(3.6)$$
This result holds provided that we do not integrate over
the singularity at the sonic point (see below).  The second equality
in equation [3.6] follows because the first integral of the
equation of motion (and therefore $f$) must vanish at the values
of density $\rho_1$ ($\ne 0$) and $\rho_2$ (by definition). Thus, the total
charge $Q$ must vanish.

Result 1 greatly limits the allowed types of wave behavior in
systems with self-gravity.  For example, one important
consequence of this ``No-Charge Property'' is that the
charge density $q(\rho)$ must vanish in the asymptotic limit
$\xi \to \infty$ for a solitary wave or soliton.  For these
waves, the mass density $\rho(\xi)$ approaches a constant
as $\xi \to \infty$.

In Paper I, we considered the case of molecular clouds with no
magnetic fields and found solutions corresponding to stationary
nonlinear waves.  For this case, the charge density is given
by $q(\rho) = \rho$ and thus the total ``charge'' $Q$ of the
stationary wave is simply the total mass contained in one
wavelength; this mass must be positive. Therefore, by Result 1,
{\it neither} soliton nor stationary wave solutions can arise
for this physical system.

The apparent contradiction between Result 1 and the stationary
wave solutions found in Paper I can be resolved by examination
of their singularity.  The stationary wave solutions of Paper I
contain a singularity, $\rho_\xi \to \infty$ as $\rho \to \rho_C$,
where $\rho_C$ satisfies $\partial p / \partial \rho = {A^2 / \rho^2}$.
Since $v= A/\rho$, the singularity is associated with the sonic point
of the fluid, i.e., where the ``effective'' sound speed of the
fluid is given by $a_{\rm eff}^2 \equiv$ $\partial p / \partial \rho$.

In the presence of the singularity, $d(f^{1/2})$ changes sign relative
to $d\rho$ across $\rho_C$, and we must rewrite equation [3.5] as
$$Q = 2{\sqrt 2} \Biggl[\int_{\rho_1} ^{\rho_C} d(f^{1/2})
-\int_{\rho_C} ^{\rho_2} d(f^{1/2})\Biggr]
\;.  \eqno(3.7)$$
The total charge is then given by
$$Q = 2{\sqrt 2} \Biggl[ 2f(\rho_C)^{1/2} -
f(\rho_2)^{1/2} - f(\rho_1)^{1/2} \Biggr] \neq 0 \, . \eqno(3.8)$$
The discontinuity that allows these solutions to avoid the
No-Charge Property necessarily produces a discontinuity in the
gravitational force $g(\xi)$, although the density profiles
of the waves are continuous.  As a result, the solutions of Paper I
for the $q(\rho)=\rho$ theory are unphysical.  In this present work,
we shall therefore require that realistic
solutions of our equations be nonsingular.

The most effective way to ensure nonsingular equations
is to limit the discussion to a mass density range
that excludes the sonic point $\rho_C$.  However, a nonsingular
class of solutions can be found by choosing the constant of
integration in equation [2.20] such that
$f(\rho_C)=0$.  For the theories considered in this paper,
this class of solutions are characterized
by a lower density bound of $\rho_1=0$ and a non-zero charge.
We will consider this class of solutions briefly
for the case of Jeans theory (see \S 4 and Appendix B).

\bigskip
\centerline{\it 3.2 Relationship Between Solitary Waves and Jeans Stability}
\medskip

In this section, we discuss the relationship between the
existence of solitary waves and the absence of Jeans instability.
This relationship arises because a soliton is in some sense
an infinite wavelength perturbation on a uniform medium.
On the other hand, the existence of a Jeans length implies
that all perturbations with a sufficiently large wavelength
will collapse. Thus, the existence of a stationary perturbation
of infinite wavelength (a soliton) is inconsistent with the
presence of a Jeans length. This relationship can be stated
more precisely as follows:

\proclaim Result 2. Suppose a physical system obeys a generalized
model equation of motion of the form [2.18] and this model equation
has solitary wave solutions. Then the system can have a uniform
density state that is Jeans {\it stable} for {\it arbitrarily large
length scales}.
\par

\noindent
Suppose we have such a system.  We can write the
equations of motion in the form [2.1], [2.2], and [2.6].
We now consider a standard Jeans-type stability analysis,
i.e., we take the unperturbed state to be $\rho = \rho_0$
= {\it constant} and $u = 0$.  (The existence of such a
uniform density state requires $q(\rho_0) = 0$; systems
which have solitary wave solutions always have such a
zero -- see the discussion of Result 3 below.)
We expand the first order quantities according to
$$f = f_0 + f_1 {\rm e}^{i(kx - \omega t)} \, , \eqno(3.9)$$
where $f_1$ is a constant. After some algebra, the dispersion
relation can be written in the form
$$\omega^ 2 = \bigl( {\partial p \over \partial \rho} \bigr)_0
k^2 - \rho_0 \bigl( {d q \over d \rho } \bigr)_0
\, , \eqno(3.10)$$
where the subscript denotes that the quantities are evaluated at
the unperturbed density. Thus, the system can be stable to
perturbations of all lengthscales provided that
$$\bigl( {d q \over d \rho } \bigr)_0 \le 0 \,  \eqno(3.11)$$
for some density $\rho = \rho_0 > 0$ such that $q(\rho_0) =$ 0.

We now show that the condition [3.11] can be realized for
systems which have solitary wave solutions. The generalized
model equation of motion can be solved for $q(\rho)$ to obtain
$$q(\rho) = - {\rm e}^{-\mu/2} {d \over d \rho}
\Bigl\{ {\rm e}^\mu {\cal F} (\rho) \Bigr\} \, , \eqno(3.12)$$
where $\cal F$ is the first integral of the equation of
motion (see equation [2.19]) and
where $\mu$ is an integrating factor,
$$\mu \equiv - 6 \log \rho + 2 \log \big[ \rho^2 {\partial p \over
\partial \rho} - A^2 \big] \, . \eqno(3.13)$$
We thus obtain an expression for $dq/d\rho$:
$${d q \over d \rho } =
- {\rm e}^{-\mu/2} {d^2 \over d \rho^2}
\Bigl\{ {\rm e}^\mu {\cal F} (\rho) \Bigr\}
+ {\rm e}^{-\mu/2} {d \over d \rho}
\Bigl\{ {\rm e}^\mu {\cal F} (\rho) \Bigr\}
{\mu \over 2} {d \mu \over d \rho}
\, . \eqno(3.14)$$
Since the system has solitary wave solutions (by hypothesis),
we know that for those solutions the quantity ${\rm e}^\mu {\cal F}$
has a double zero with a positive second derivative at some density
$\rho = \rho_0$.  Equation [3.14] shows that for $\rho = \rho_0$,
the quantity $dq/d\rho$ is negative.  Thus, by equation [3.11],
a Jeans stable configuration can arise for
this physical system.

We note that the converse of Result 2 is not true.  Physical
systems of this type can be Jeans stable to perturbations of
all wavelengths and still not have soliton solutions.  Thus,
the requirement of Jeans stable configurations represents a
necessary condition (and not a sufficient condition) for the
existence of solitons. We consider other related conditions
in the following section.

\bigskip
\centerline{\it 3.3 Necessary and Sufficient Conditions}
\centerline{\it for the Existence of Stationary Waves and Solitons}
\medskip

In this section, we consider the conditions required for one dimensional
systems to exhibit solitary waves and stationary waves.  The
question of whether or not solitary wave solutions exist is
fundamental to the study of nonlinear dynamics and cannot, in
general, be answered in a definitive manner. In this paper,
we have shown that a fairly large class of physical systems can
be modeled using a theory with a charge density $q(\rho)$.  The
question can then be posed as follows: What properties must the
charge density $q(\rho)$ have in order for solitary wave solutions
to exist and what properties are required for stationary waves to
exist? Fortunately, we can provide a partial answer to this question.

We first find a {\it necessary} condition on the charge density.
In order for solitary wave solutions to exist, the first integral
$\cal F$ of the equation of motion (see equation [2.21]) must be
positive between a double zero and a ordinary zero (see \S 2).
Using the solution in the form of equation [2.19], we can write
$${\cal F} (\rho) = \rho^6 \Bigl[ \rho^2 \dpdrho - A^2
\Bigr]^{-2} f(\rho)  \, , \eqno(3.15)$$
where $f(\rho)$ is defined by equation [2.20] and depends
on the form of the charge density.  We restrict our discussion
to waves which avoid the singularity at the sonic point.
We also restrict our discussion to waves with
finite velocity; this second restriction eliminates the point
$\rho = 0$ as a candidate for the double zero (if we let
$\rho \to 0$, we get $v =A/\rho \to \infty$ by the continuity
equation).  Thus, the required double zero of $\cal F$ must
occur at a double zero of $f(\rho)$.  Furthermore, $f(\rho)$ must
be a local minimum at the double zero.  In addition, in order for the
second (ordinary) zero of $\cal F$ to exist, the function $f(\rho)$
must also have a local maximum (at some other density).  By definition,
the derivative of $f(\rho)$ is given by
$${df \over d \rho}  = {q(\rho)\over \rho} \,
\bigl[ {A^2 \over \rho^2}  - \dpdrho \bigl] \, .
\eqno(3.16)$$
Since we must avoid the singularity at the sonic point, both the
required minimum and the maximum of $f (\rho)$ must occur where
the charge density $q(\rho)$ vanishes.  We denote the location of
the minimum as $\rho_1$ and the maximum as $\rho_M$.  In order for
the point $\rho_1$ to be a minimum, $dq/d \rho$ must have the
correct sign, which depends on the
sign of the quantity in brackets in equation [3.16].
In other words, the sign depends on whether we are considering
purely subsonic or purely supersonic waves.  For subsonic waves
we require $dq/d \rho < 0$ at $\rho_1$ and $dq/d \rho > 0$
at $\rho_M$.  For supersonic waves, the sign requirements
are reversed.  These arguments can be
summarized as the following necessary condition on the charge
density $q(\rho)$ for the existence of solitary waves:

\proclaim Result 3. In order for solitary wave solutions to
exist for the class of theories considered in this paper, the
charge density $q(\rho)$ must have (at least) two zeroes
$\rho_1$ and $\rho_M$.  For subsonic waves, $q(\rho)$ must
be negative between the two zeroes; for supersonic waves,
$q(\rho)$ must be positive between the zeroes. (This result
applies only to nonsingular solutions where $\rho_C$ does
not lie between $\rho_1$ and $\rho_M$).
\par

Result 3 is potentially very powerful as a test to see if
solitary wave solutions exist.  For example, many possible
charge density functions do not have two zeroes and thus
solitary wave behavior can be easily ruled out.
We note, however, that Result 3 is not sufficient to
guarantee the existence of solitary waves.  We must place
an additional constraint on the charge density to make sure that
the second (ordinary) zero of the function $f(\rho)$ exists.
Given the two zeroes of $q(\rho)$, the possible behavior of
the function $f(\rho)$ is sketched in Figure 1.  The first
possibility is that $f$ has a second ordinary zero (solid
curve in Figure 1) and thus solitary wave solutions exist.
If the function
$f$ has another minimum at some density $\rho_3 > \rho_M$ and
hence turns up (as shown by the short dashed curve in Figure 1),
then we can always choose the constant of integration $\beta$
to make $\rho_3$ a double zero of $f$.  We thus obtain a
{\it depression soliton} or {\it void} solution.
Only for the special case where $f(\rho)$ approaches
a constant asymptotically (long dashed curve in Figure 1)
does the system fail to have solitary wave solutions.

For the usual case of purely subsonic waves, the existence of
an ordinary second zero of $f$ is guaranteed provided that the
function $f \to - \infty$ as $\rho \to \infty$.  In the limit
$\rho \to \infty$,
$${df \over d\rho } = - {1 \over \rho}
{\partial p \over \partial \rho}
q(\rho) \, , \eqno(3.17)$$
where we have assumed that the equation of state is well behaved
(i.e., the pressure increases with increasing density).
Thus, the requirement that
$$\lim_{\rho \to \infty}
{1 \over \rho} {\partial p \over \partial \rho} q (\rho)
> 0 \, \eqno(3.18)$$
is sufficient, but not necessary, to ensure the existence
of the second zero of $f(\rho)$ for the case of subsonic waves.
For most of the examples considered in this paper, $q \to \rho$
and $\partial p / \partial \rho$ $\to$ 1 in the limit
$\rho \to \infty$ (for the case of a magnetic pressure with
$p \sim \rho^2$, $\partial p / \partial \rho$ $\to \infty$);
thus, the constraint [3.18] is always
satisfied for these theories.

We now consider the simpler case of ordinary stationary waves.
The required conditions on the charge density $q(\rho)$
for the existence of these waves can be stated as follows:

\proclaim Result 4. In order for stationary wave
solutions to exist for the class of theories considered in this
paper, the charge density $q(\rho)$ must have (at least) one zero
$\rho_M$.  For subsonic waves, $dq/d\rho > 0$ at $\rho_M$ and this
zero must occur at a density larger than the sonic density $\rho_C$
(the density of singularity).  Similarly, for supersonic waves,
we must have $dq/d\rho < 0$ at $\rho_M < \rho_C$.
\par

The conditions outlined in Result 4 are both necessary
and sufficient for the existence of stationary waves.
This result is straightforward to understand.  In order
for such waves to exist, the function $f(\rho)$ must
have a maximum and two zeroes $\rho_1 < \rho_2$ such
that $\rho_C < \rho_1$ for subsonic waves and
$\rho_C > \rho_2$ for supersonic waves.
The zero of $q(\rho)$ is required to provide a critical
point for $f(\rho)$ -- see equation [3.16].  The sign
requirement on $dq/d\rho$ ensures that
the critical point is a maximum.  The third condition,
that the critical point $\rho_M$ be larger than the
sonic point for subsonic waves and smaller than the
sonic point for supersonic waves, is required so that
the singularity does not lie in the range of densities
in the wave profile.

\bigskip
\centerline{\bf 4. NONLINEAR WAVES AND SOLITONS IN JEANS THEORY}
\medskip

In this section, we explicitly consider the case of fluid systems
with a charge density of the form $q = \rho - \rho_0$.
As discussed above, this choice for the charge density
is equivalent to adopting the original approximation of
Jeans or to considering a uniformly rotating cloud (see \S 2.2).
This system is also the simplest modification of the basic
fluid equations that allows for the existence of stationary waves.
Application of the No-Charge Property to this system requires
$\rho_0$ to be the average density for stationary wave solutions.

We can analyze the possible behavior of this system using
the results of \S 3.  The charge
density $q(\rho)$ has only a single zero (at $\rho = \rho_0$)
and hence Result 3 shows that no solitary wave solutions
exist (except for solutions with vanishing density as $\xi \to \infty$;
see Appendix B).  In addition, $dq/d\rho = 1 > 0$ everywhere
and Result 4 implies that subsonic stationary nonlinear waves
are possible provided that the singularity at the sonic point
can be removed from the wave profile.  This requirement
implies that $\rho_0 > \rho_C$ for subsonic waves and
can be written as a constraint on the Mach number $A$:
$$A^2 < \rho_0^2 \Bigl[ {\partial p \over \partial \rho}
\Bigr]_{\rho_0} \, . \eqno(4.1)$$
This condition thus defines a maximum value of the Mach
number $A$ for nonlinear stationary waves.

In order to study the properties of the stationary waves,
we must specify the equation of state and find the function
$f(\rho)$ defined by equation [2.20].
For the sake of definiteness, we take the pressure to have
the form given by equation [2.4b].  For this equation of state,
the integral in equation [2.20] can be evaluated to obtain
$$f(\rho)= \beta - \rho - (\kappa - \rho_0) \log(\rho)-
{{A^2+\kappa\rho_0}\over \rho} + {A^2\rho_0\over 2\rho^2} \, ,
\eqno(4.2)$$
where $\beta$ is an integration constant.  The function
$f(\rho)$ does allow for wave solutions; a schematic
is shown in Figure 2.

Without loss of generality, we can choose $\rho_0=1$ (see Appendix A).
This choice leaves us with a two parameter family of wave solutions.
The parameter $A$ sets the velocity of the wave and also determines
the value of $\rho_C$.  The parameter $A$ can vary from zero up to
a maximum given by the condition [4.1].
For a given value of $A$, the constant of
integration $\beta$ merely slides $f(\rho)$ in the vertical direction
(see Figure 2) and thus determines the wave amplitude, which can vary
from zero (when $\rho_1=\rho_2=\rho_0$) to a maximum value
(when $\rho_1=\rho_C$).  In general, the range of possible amplitudes
decreases with increasing $A$.  A typical example of a nonlinear
stationary wave in Jeans theory is shown in Figure 3. Note that
the wavelength is $\lambda \sim 20$ for this example; in
physical units (scaled to the Taurus cloud -- see the discussion
following equation [2.5]), the wavelength is $\sim$10 pc.

Given the solutions for nonlinear stationary waves discussed above,
we can determine the allowed wavelengths for nonlinear waves as a
function of amplitude.  For specified values of the constants
$A$ and $\beta$, the wavelength of a stationary wave is given by
$$ \lambda = 2 \int_{\rho_1}^{\rho_2}\,
\left({d\rho\over d\xi}\right)^{-1}\, d\rho  \, , \eqno(4.3)$$
an integral which unfortunately must be done numerically.
In Figure 4 we plot the wavelength as a function of wave amplitude
$\delta\rho/\rho \equiv ( \rho_2- \rho_0 )/\rho_0$
for two extreme cases.   The first case has $A=0$ and
corresponds to the ``Jeans'' length generalized for
nonlinear waves; note that we recover
the linear Jeans length for small amplitude.  In fact, the Jeans
length is shown to change very little over a large range of
amplitudes.  The opposite limit corresponds to the ``shock''
limit; the constraint that all velocities must remain below
the effective sound speed $a_{\rm eff}$
gives a minimum possible wavelength for a wave of a given
amplitude.  This minimum wavelength approaches the ``Jeans''
length rapidly with increasing amplitude, so that for
$\delta\rho/\rho\gtwid 1$ only a narrow range of wavelengths is
possible.

Nonlinear waves are characterized by relatively narrow,
high density peaks.  While the wavelength corresponds to
the distance between peaks, the width of the peaks is also
an interesting quantity.  In order to obtain a measure of
the width of the wave crests, we define the quantity $\lambda_p$
to be the length of the region of the wave profile where
$\rho > \rho_0$, i.e.,
$$ \lambda_p \equiv 2 \int_{\rho_0}^{\rho_2}\,
\left({d\rho\over d\xi}\right)^{-1}\,
d\rho\, . \eqno(4.4)$$
In Figure 5, the width $\lambda_p$ is plotted as a function of wave
amplitude for both the ``Jeans'' and the ``shock'' limits described
above.  Figure 5 clearly shows that for $\delta\rho/\rho\gtwid 1 $, the
range of allowed values of $\lambda_p$ is very tightly constrained.
In fact, the width $\lambda_p$ has almost a single value as
a function of $\delta\rho/\rho$.

These results have a relatively simple interpretation.
For stationary waves, a balance must exist between
dispersion, which acts to spread the wave, and nonlinear
effects, which act to steepen the wave and cause it to shock.
For the case at hand, the dispersion is supplied entirely by the
self-gravitational interaction. (Recall that in the linear Jeans
analysis the dispersion relation is given by
$\omega^2= a_{\rm eff}^2 k^2- 4\pi G\rho_0$.)
For waves with $\delta\rho/\rho\gtwid 1$, nonlinear effects will
be large.  To balance these large nonlinear effects, the effects
of self-gravity must also be large, hence the observation that the
natural length scale for nonlinear stationary waves is of order the
Jeans length.  The narrow range of allowed stationary waves for
large amplitudes $\delta\rho/\rho\gtwid 1$ is thus a consequence
of the fact that the amount of self-gravity needed to balance the
nonlinearity is comparable to that required to cause the wave to
collapse.

For completeness we note that a class of solitary wave solutions
exists for Jeans theory.  In these solutions, the singularity is
removed by choosing the integration constant $\beta$ appropriately
(see Appendix B for further discussion).  An example of such a
solitary wave is shown in Figure 6.

\bigskip
\bigskip
\centerline{\bf 5. NONLINEAR WAVES AND SOLITONS}
\centerline{\bf IN CLOUDS WITH MAGNETIC FIELDS}
\medskip

In this section we consider the effects of magnetic fields, which
are expected to play an important role in the dynamics of molecular
clouds (see, e.g., Shu et al. 1987). In Paper I, we derived a
model equation which incorporates the effects of magnetic fields
in the equations of motion and produces a charge density theory
(see the review in \S 2.3 and equation [2.10]).  This model equation
assumes that the largely neutral fluid is coupled to the magnetic
fields through the ionic component as an intermediary.  In order to
obtain a simple form for the equation of motion, a dissipative term
was dropped from consideration.   Since this approximation has been
discussed at some length in Paper I, it will not be considered here.

Here we adopt an alternate approach; we assume that the magnetic
field is strongly coupled to the neutral fluid.  This assumption is
valid for large lengthscales in molecular clouds (scales larger
than those associated with dense cores), since the magnetic
diffusion time is long compared to the magnetic sound crossing
time (see Shu 1992).  In fact, the existence of stationary wave
solutions requires the absence of diffusive effects (recall that
in Paper I we also ignored a diffusion term).

With the neutrals strongly coupled to the magnetic field, the ions
do not explicitly enter in the dynamics (the ions act only as an
intermediate coupling agent).  The equation of motion for the
neutral component becomes
$${\partial u \over \partial t} + u
{\partial u \over \partial x} + {1 \over \rho}
{\partial p \over \partial x} +
{\partial \psi \over \partial x} = {1\over\rho}
\left[\left(\nabla\times {\bf B}\right)\times {\bf B}\right]
\cdot {\hat x} , \eqno(5.1)$$
where we have considered only one spatial dimension and where
$B$ is written in dimensionless form $B\to B/ (4\pi a_s^2\rho_R)^{1/2}$.
Observations of magnetic field strengths in molecular clouds (e.g.,
Goodman et al. 1989; Myers \& Goodman 1988) indicate that $B$ =
10 -- 30 $\mu$G (with these values of field strength, the
dimensionless version of $B$ will be of order unity).
The equation of motion [5.1] must be supplemented by an
equation of motion for the magnetic field itself,
$${\partial {\bf B}\over\partial t} =
\nabla\times(u\hat x\times {\bf B}) \; . \eqno(5.2)$$

Given the form of the magnetic force term, we see that the
one-dimensional formalism is consistent as long as the magnetic
tension does not result in a $y$-directed force (since no other
forces exist to balance such a term).  The magnetic field must
also satisfy the usual divergenceless (no-monopole) condition
$$\nabla\cdot {\bf B} = 0\, .  \eqno(5.3)$$
These two requirements imply that the magnetic field cannot
have components both parallel and perpendicular to $\hat x$.
Since a parallel component has no effect on the neutral
dynamics, we consider a magnetic field of the form
$${\bf B} = B\hat y \, . \eqno(5.4)$$
Equations [5.1] and [5.2] may therefore be expressed as
$${\partial u \over \partial t} + u
{\partial u \over \partial x} + {1 \over \rho}
{\partial \over \partial x} \left(p+{B^2\over 2}\right)+
{\partial \psi \over \partial x} = 0, \eqno(5.5)$$
and
$${\partial B  \over \partial t } + {\partial \over \partial x}
\left(u B\right)=0  . \eqno(5.6)$$

In the stationary wave approximation, the equation of motion [5.6]
for the magnetic field can be integrated to obtain the relation
$$B \, v = constant \, . \eqno(5.7)$$
This relation, along with the solution [2.17] to the continuity
equation, gives the expected flux-freezing condition for a
one-dimensional system,
$${B\over\rho} = \alpha = constant\;.\eqno(5.8)$$
In typical molecular clouds, the magnetic contribution to the
pressure is comparable to the ``turbulent'' contribution and
hence we expect $\alpha^2 \sim \kappa$.
We can use the flux freezing condition [5.8] in the equation of
motion [5.5] to eliminate the magnetic field.  The resulting
equation of motion (including magnetic effects) is related
to the previous one (equation [2.19]) through the transformation
$p \to p + \alpha^2 \rho^2 /2$.  In other words, we have simply
added a magnetic pressure term to the nonmagnetic case.  This new
pressure is given by
$$p (\rho) = \rho + \kappa\hbox{log}(\rho) +
\alpha^2\rho^2/2 \; , \eqno(5.9)$$
and we can proceed as in \S 4.  For the values expected in
molecular clouds, the second and third terms in equation [5.9]
are comparable in size.  In other words, the ``turbulent''
contribution to the pressure is comparable to the magnetic
pressure.

Here, we take the charge density $q(\rho)$ to be the same as
in the previous section.  The results of \S 3 thus imply that
the system cannot have solitary wave solutions, but can have
subsonic stationary wave solutions.  To study these waves,
we find the function $f (\rho)$ which becomes
$$f(\rho)= \beta - {\alpha^2\over 2}\rho^2+
\left(\alpha^2\rho_0-1\right)
\rho - (\kappa - \rho_0) \log(\rho)-
{{A^2+\kappa\rho_0}\over \rho} + {A^2\rho_0\over 2\rho^2}
\eqno(5.10)$$
where $\beta$ is an integration constant.  As we found in \S 4,
the function $f(\rho)$ is positive for small $\rho$ and becomes
negative as $\rho \rightarrow \infty$.  For this form of $f$,
the wave solutions are analogous to those found in \S 4 and hence
the discussion given there also applies to this present case.

Figure 7 presents the density profile found by integrating
equation [2.19] for the pressure given by equation [5.9].  The
parameters used are representative of cloud environments.
Comparison of Figure 7 with Figure 3 shows that a strong magnetic
field is capable of sustaining a broader wave structure.  This
result is expected since the added magnetic pressure provides
additional support against self-gravity and because the
equation of state [5.9] has a stiffer component.
As in the case of pure Jeans theory (\S 4), nonlinear waves
can only exist in a fairly narrow range of wavelengths.
The allowed range of wavelengths for molecular clouds
with magnetic pressure is shown in Figure 8.  Notice that
the maximum allowed wavelength varies more rapidly with
amplitude than in the case of Figure 4.

\bigskip
\centerline{\bf 6. NONLINEAR WAVES AND SOLITONS WITH YUKAWA POTENTIALS}
\medskip

In this section, we present a class of model equations with
charge densities arising from modifications of the long range
forces of gravity.  In particular, we consider theories in
which gravity is modeled with a Yukawa potential.
As derived in \S 2.4, this approximation leads to a
charge density theory with $q(\rho)$ of the form
$$q(\rho) = \rho - m^2 \bigl[ h(\rho) +
{A^2 \over 2 \rho^2} - E \bigr] \, . \eqno(6.1)$$
With this form for the charge density, the function
$f(\rho)$ can be easily found:
$$f(\rho) = \beta - p(\rho) - {A^2 \over \rho} + {1 \over 2}
m^2 \bigl[ h(\rho) + {A^2 \over 2 \rho^2} - E \bigr]^2 \, .
\eqno(6.2)$$
Notice that for the equation of state [2.4b], the charge
density $q(\rho)$ is of cubic order and has either one or
three real zeroes; thus, solitary wave behavior is possible.

We now show that all stationary waves in this theory must
propagate at subsonic speeds.  Result 4 shows that supersonic
stationary waves must have $q = 0$ and $dq/d\rho < 0$
at some density $\rho_M < \rho_C$.  The first of these
conditions ($dq/d\rho < 0$) implies that
$${\partial p \over \partial \rho}  - {A^2 \over \rho^2} >
{\rho \over m^2 } \, .  \eqno(6.3{\rm a})$$
On the other hand, the requirement that the wave is
supersonic ($\rho_M < \rho_C$) implies that
$${\partial p \over \partial \rho}  - {A^2 \over \rho^2}
< 0 \, , \eqno(6.3{\rm b})$$
where both conditions [6.3a] and [6.3b] are to be
evaluated at the zero of $q$.  Since both of these
conditions cannot be simultaneously satisfied,
no supersonic wave solutions exist for this theory.
In other words, all physical stationary wave solutions
must propagate at subsonic velocities.

We now consider solitons in this theory.  Result 3 shows
that soliton solutions require the charge density $q (\rho)$
to have at least two zeroes; for subsonic waves, $q$ must be negative
between two zeroes.  Suppose $q(\rho)$ has a zero at $\rho_S$.
We can always choose the constant of integration $\beta$ to make
the point $\rho_S$ a double zero of $\cal F$.  In this case,
$\rho_S$ represents the fluid density
far from the soliton.  Without loss of generality, we can
rescale the variables so that $\rho_S$ = 1 (see Appendix A).
In order for the charge density to be negative between $\rho_S$
and a second zero of $q$,  we must have
$${d q \over d \rho} = 1 - {m^2 \over \rho_S}
\Bigl[ {\partial p \over \partial \rho} - {A^2 \over \rho^2}
\Big]_{\rho_S}  \, < \,  0 \, , \eqno(6.4)$$
where the term in square brackets is to be evaluated
at $\rho = \rho_S = 1$.  Thus, a necessary condition for
the existence of soliton solutions is that
$$m^2 > \Biggl[ {\partial p \over \partial \rho} \Big|_1 - A^2
\Biggr]^{-1}  \, .\eqno(6.5)$$
This condition has two important consequences.  First, we note
that for the theory to have solitons at all, the quantity
$m^2$ must satisfy the condition
$$m^2 > \Bigl[ \, \, {\partial p \over \partial \rho} \Big|_1
\, \, \Bigr]^{-1} \, .\eqno(6.6)$$
This constraint is equivalent to the condition that the theory
has no Jeans length, as obtained from the linear dispersion
relation (see also Result 2).  Second, if the value of $m^2$ satisfies
equation [6.6], then the Mach number $A$ must satisfy the constraint
$$A^2 < \left[ {\partial p \over \partial \rho} \Big|_1
- {1\over m^2} \right] \, . \eqno(6.7)$$
Since $A/\rho_S$ is the velocity of the soliton in the rest
frame of the fluid, this constraint implies a maximum soliton
velocity for a given value of $m$.

The conditions outlined above are necessary for the existence
of solitons.  For equations of state for which $f \to - \infty$
as $\rho \to \infty$, these conditions are also sufficient.
Notice that this applies for our usual choice of
equation of state [2.4b].

Thus far, our discussion has been limited to compression solitons,
i.e., waves in which the density of the soliton is enhanced over
that of the background.  However, under certain
conditions, depression solitons (voids) can exist as well. This
type of wave behavior can occur if the function $f$ drops below
zero for $\rho_C < \rho < 1$.  For the equation of state [2.4b],
we can show that a necessary and sufficient condition for the
existence of depression solitons is that $f(\rho_C) < 0$.  After
some straightforward algebra, this requirement can be written
as a condition on the parameter $m$,
$$m^2 < {2\left[ p(\rho_C) - p(1) - A^2 (1-\rho_C)^2 / 2\rho_C^2
- h(\rho_C) + h(1) \right] \over \left[ h(\rho_C) - h(1)+ A^2
(1 - \rho_C^2) / 2\rho_C^2 \right]^2} \, . \eqno(6.8)$$
In Figure 9, we show wave profiles for both a standard (compression)
soliton and a depression (or void) soliton.

The constraints derived above define regions of allowed parameter
space in the $(m,A)$ plane.  In Figure 10, we plot these
constraints for the equation of state given in equation [2.4b]
with $\kappa$ = 10.  The vertical dashed line on the right of
the figure corresponds to the sonic point; all wave solutions
in this theory must be subsonic and hence $A$ must lie to the
left of this line.  The constraints of equations [6.5] and [6.8]
are shown as solid curves.  In region I, ordinary (compression)
soliton solutions are allowed; in region II, both compression and
depression (void) solitons are allowed.  If the constraint [6.5]
is not satisfied, then no solitary waves exist (this case
corresponds to region III of Figure 10).  However, the theory
does allow nonlinear waves of various amplitudes, which are
determined by varying $\beta$.  In analogy with the compression
and depression solitons described above, these nonlinear waves
can have profiles with narrow valleys as well as the usual
profiles which are narrowly peaked.

The inclusion of a magnetic pressure term $\alpha^2 \rho^2 /2$
(see \S 5) can significantly change the results described
above.   For this case, $h \to \alpha^2 \rho$ in the limit
$\rho \to \infty$ and thus
$f \to (m^2 \alpha^4  - \alpha^2) \rho^2/2$.
For $m^2 \alpha^2 > 1$, the function $f \to \infty$
as $\rho \to \infty$ in constrast to the behavior discussed
above (see the discussion following equation [6.2]).
For this pressure law, compression solitary wave solutions
do not exist when $m^2 \alpha^2 > 1$.

\bigskip
\centerline{\bf 7. APPROXIMATION OF TWO-DIMENSIONAL THEORY}
\medskip

In this section we derive one dimensional model equations by
approximating two dimensional fluid systems.  The goal of this
procedure is to obtain a qualitative understanding of wave motions
in higher dimensions while retaining the mathematical simplicity
of the one dimensional theory.

We begin with the equations of motion for a two dimensional
fluid.  We consider stationary wave solutions, so we define
the variable $\xi = x - v_0 t$ and let $v = u - v_0$
be the velocity along the $\hat x$ direction as before.
We let $w$ denote the velocity along the $\hat y$ direction.
The continuity equation in this stationary wave approximation
becomes
$$\partial_\xi (\rho v) + \partial_y (\rho w ) = 0 \, .
\eqno(7.1)$$
The two components of the force equation become
$$v v_\xi + w v_y + h_\xi + \psi_\xi = 0 \, , \eqno(7.2)$$
$$v w_\xi + w w_y + h_y + \psi_y = 0 \, , \eqno(7.3)$$
and the Poisson equation for the gravitational potential is
$$\psi_{\xi \xi} + \psi_{y y} = q_0 (\rho) \, , \eqno(7.4)$$
where we have left open the possibility of a nontrivial
charge density $q_0 (\rho)$.

We now consider a plane wave traveling in the $\hat x$ direction.
However, we want the plane wave to have finite extent in the
perpendicular ($\hat y$) direction.  As a first approximation,
we assume that the flow velocity in the transverse direction is
much smaller than that in the direction of propagation and
set $w = 0$.  We are thus considering a wave structure with
a more or less fixed profile in the transverse direction.
In this case, the $\hat y$ component of the force equation [7.3]
reduces to the hydrostatic condition $h_y = - \psi_y$, and the
Poisson equation can be written in the suggestive form
$$\psi_{\xi \xi} = q_0 (\rho) + h_{y y} \, . \eqno(7.5)$$

To make further progress, we must specify the equation of state
(in order to specify the form of the enthalpy), the charge density
$q_0$, and the transverse profile of the wave.  We note that
the equation of state in the transverse direction need not
be the same as that in the direction of wave propagation;
in other words, the pressure need not be isotropic.  For
example, magnetic fields can provide support in the transverse
direction. For simplicity, however, we consider an equation of
state of the form [2.4b] and the enthalpy becomes
$$h(\rho) = \log \rho - \kappa / \rho \, . \eqno(7.6)$$
In order to isolate any possible pseudo-two-dimensional effects,
we adopt a trivial form $q_0 = \rho$ for the original charge
density.  If we assume that the wave structure is a localized
lump in the transverse direction, we can take the density
profile to have the separable form
$$\rho = \rho( \xi) \exp [- \ypro^2 y^2 ] \, . \eqno(7.7)$$
This form is chosen because it is symmetric about the $y=0$ plane
and decreases with $y$.  The parameter $\ypro$ is like a wavenumber
that sets the scale of the structure in the transverse direction.
Using this ansatz for the density profile, we obtain a particularly
simple form for the term $h_{y y}$, namely,
$$h_{y y} = - 2 \ypro^2 (1 + \kappa/\rho) - 4 \ypro^4 y^2
\kappa/\rho \, . \eqno(7.8)$$
Because the equation of motion in nonlinear, we cannot
self-consistently assume a profile of the form [7.7]
for all values of $y$.  However, near the center of the wave
profile (i.e., near $y=0$), we can neglect the $y$ dependence
everywhere except for keeping the first term in equation [7.8].
We thus obtain the Poisson equation in the form
$$\psi_{\xi \xi} = \rho - 2 \ypro^2 (1 + \kappa/\rho)
\equiv q(\rho) \, , \eqno(7.9)$$
where in the second equality we have defined the (total)
effective charge density including two dimensional effects.
We note that the form [7.9] is not sensitive to the
particular form of the wave profile in the transverse
direction; we simply require that the profile is even
in $y$ and decreases on a lengthscale given by $\mu^{-1}$.
We stress that we have {\it not} solved the full two dimensional
problem; we have simply assumed a reasonable structure in the
transverse direction.  The physical interpretation of this
approximation is that if the wave structure has finite spatial
extent in the transverse direction, then the gravitational field
falls off with increasing distance (in the $\hat x$ direction).
For the particular simple form of the transverse wave structure
used here, we obtain yet another charge density theory, which
can be solved using the methods developed in this paper.

Given the form of the charge density (equation [7.9]), we see
immediately that this theory does not allow for solitary wave
solutions (because the charge density $q(\rho)$ has only one
zero with positive density -- see Result 3).
We also see immediately that $dq/d\rho$ is always positive;
thus, by Result 4, only subsonic nonlinear stationary waves
are allowed.  In order to ensure that physically relevant
wave solutions exist, we still must show that the sonic
point can be removed from the range of densities of the
wave profile.  This final requirement can be met if the
zero of $q(\rho)$ occurs at a density larger than that of
the sonic point, i.e., provided that
$$\ypro^2 + \ypro \big[ \ypro^2 + 2 \kappa \big]^{1/2} >
- {\kappa \over 2} + {1 \over 2}
\big[ \kappa^2 + 4 A^2 \big]^{1/2} \, . \eqno(7.10)$$
This condition can clearly be satisfied if the parameter
$\ypro$ is large enough, i.e., if the density profile
in the transverse direction falls off sufficiently rapidly.
In order to obtain a better feeling for how restrictive this
condition actually is, we consider the limit $A \ll \kappa$
(typically, $A \sim 1$ and $\kappa \sim 10$, so this limit
is quite reasonable for molecular clouds).  In this case, the
constraint [7.10] reduces to the form
$$\ypro^2 > {A^4 \over 2 \kappa^3 } \, , \eqno(7.11)$$
where we have kept terms to leading order in $A^2/\kappa^2$.
Using the values $A \sim 1$ and $\kappa \sim 10$,
we find that $\ypro > 0.022$ is sufficient
to allow for stationary wave solutions to exist.
In other words, the length scale of the density fall-off in
the transverse direction can be 45 times the thermal Jeans length
and still allow for stationary waves (see equation [7.7]).

Before leaving this section, we note that the Yukawa theory
of the previous section can be recovered by using a
pseudo-two-dimensional argument similiar to that given
above. We write the Poisson equation as
$$\psi_{\xi \xi} = \rho - \psi_{y y} \, \eqno(7.12)$$
and then expand the equation about the plane $y=0$.
If we assume that the potential $\psi$ is separable
near $y=0$, we can write
$$\psi_{y y} = - \mu^2 \psi(\xi) \, . \eqno(7.13)$$
Combining equations [7.12] and [7.13], we obtain the
Poisson equation for a Yukawa theory (see equation [2.11]).

\bigskip
\centerline{\bf 8. DISCUSSION}
\medskip

\medskip
\centerline{\it 8.1 Summary of Results}
\medskip

In this paper we have developed further the theory of wave motions
in self-gravitating astrophysical fluids.  Although many of the
results are general, the application to the fluid dynamics of
molecular clouds is our primary motivation. Our results can be
summarized as follows:

\item{[1]}  We have introduced the concept of ``charge density''
for the study of self-gravitating fluid systems (see also AFW).
In this formalism, the density is replaced by a
``charge density'' $q(\rho)$ on the right hand side
of the Poisson equation (the continuity equation and the
force equation remain as usual).  We have shown that a large class
of physical systems can be modeled with a charge density; we can
thus include a wide range of physical effects while retaining a
simple semi-analytic theory.

\item{[2]} We have proven a ``No-Charge Property'' (Result 1)
which shows that no solitons or stationary waves can exist in
one-dimensional self-gravitating fluids unless the total charge
vanishes (where the total charge is the integral of the charge
density over one wavelength).  This requirement greatly
constrains the types of model equations that
allow for stationary wave behavior.

\item{[3]} We have shown that in order for a physical system to
exhibit solitary wave behavior, the system must also be capable
of having a configuration which is stable to gravitational perturbations
{\it of arbitrarily large wavelengths} (see Result 2).  We have
thus discovered a fundamental relationship between gravitational
stability and the existence of solitary waves.

\item{[4]} We have found constraints on the form of the charge
density $q(\rho)$ for the existence of solitary waves and ordinary
stationary waves (see \S 3.3, Results 3 and 4).  These conditions
allow us to determine (or at least constrain) the possible types
of wave behavior for any charge density theory {\it without
having to solve the equations of motion}.

\item{[5]} The original Jeans analysis provides the simplest
nontrivial theory with a charge density and has
$q(\rho) = \rho - \rho_0$.  For this theory, we have studied
the propagation of nonlinear waves in one dimension and
found a class of wave solutions.  In spite of this theoretical
idealization, these waves are probably more physical than the
nonlinear wave solutions found in Paper I.

\item{[6]} Using the charge density formulation of the problem,
we have performed a nonlinear Jeans analysis for molecular clouds.
We find that the Jeans length is a slowly increasing function
of the wave amplitude $\delta \rho/\rho$.  The wavelengths for
stationary waves must be less than this Jeans length.
In addition, we find that waves with sufficiently small
wavelengths tend to shock and dissipate.  As the wave
amplitude $\delta \rho/\rho$ becomes larger, nonlinear waves
become confined to a rather narrow range of wavelengths.
These clouds thus select out a particular lengthscale as
a function of density contrast (see Figures 4 and 5).

\item{[7]} The effects of magnetic fields can be incorporated
into this theory in two conceptually different ways.  In Paper I,
we derived a model equation which takes into account the relative
drift between the neutral and ionic species; this model equation,
which omits a dissipative term, allows for a wide variety of
nonlinear wave behavior including solitary waves and topological
solitons.  In this paper, we have studied the other extreme --
flux freezing.  In this case, the magnetic field simply adds another
component to the pressure.  As in the case of Jeans theory (see
item [6]), nonlinear waves are confined to a narrow range of
allowed wavelengths (see Figure 8).

\item{[8]} We have studied theories with Yukawa potentials
which allow the graviational force to fall off with distance while
retaining a one-dimensional formulation of the problem.  This
theory contains both stationary waves and solitary waves.
In addition, depression solitons or voids are allowed.

\item{[9]} Using a two-dimensional theory as a starting point,
we have derived a model equation which takes the form of a
one-dimensional charge density theory.  We can thus heuristically
take into account two-dimensional effects while retaining a
one-dimensional formulation.  This particular model does not
allow solitary wave solutions, but does allow stationary waves.

We have thus introduced a new class of ``charge density theories''
for the study of wave motions in self-gravitating fluids.  We
have proven general results (Results 1 -- 4) which provide
powerful constraints on the allowed types of wave behavior
in these systems.  Finally, we have argued that many physical
systems can be modeled with charge density theories and we have
studied several examples relevant to molecular clouds.

\medskip
\centerline{\it 8.2 Applications and Comparison with Observations}
\medskip

Many of the results of this paper are general and can be applied to
any self-gravitating fluid system (and other similar systems with
long range forces; see AFW).  However, one primary motivation for
this work is to understand the formation of substructure in molecular
clouds.  These cloud systems are self-gravitating (see, e.g., the
reviews of Blitz 1993; Shu et al. 1987) and exhibit highly
nonlinear structures (e.g., Myers 1991; Houlahan \& Scalo 1992;
Wood, Myers, \& Daugherty 1993; Wiseman \& Adams 1993; Blitz 1993;
Adams 1992; de Geus, Bronfman, \& Thaddeus 1990).
In addition, these clouds exhibit a wide range of velocity structure
as measured by line of sight velocity variations and by varying
line-widths.  These observations indicate cloud motions which are
faster than the thermal sound speed but generally slower than the
Alfv\'en speed.  In the models of this paper, the fluid speeds are
determined by the Mach number $A$ which is normalized to the thermal
sound speed.  Thus, for applications to molecular clouds, $A$
should lie in the approximate range
$1 < A < (1 + \kappa)^{1/2} \sim 3$.

The first obvious signature of wave motions in molecular clouds
is periodic or nearly periodic structures.  Several examples
can be found in the existing literature.  For example, the star
forming region NGC 6334 contains five regularly spaced clumps,
as deduced from the 69 $\mu$m continuum map of the region (see
McBreen et al. 1979); the inferred wavelength is $\sim$3 pc.
Such behavior is not rare.  A survey of 23 globular filaments
(defined to be filamentary dark clouds which exhibit
condensations) shows that ``the most striking similarity among
all of the globular filaments is the regularity of their
segmentation'' (Schneider \& Elmegreen 1979). This survey
also shows that the characteristic spacing of the condensations
along the filaments is about three times the width of the filament.
Another example of possible wave motion is provided by the Taurus
Molecular Cloud complex, which shows evidence for a periodicity in
the velocity of 21cm observations of self-absorption (Shuter, Dickman,
\& Klatt 1987). These authors interpret their data as velocity
waves with a peak to peak amplitude of $\sim$3 km/s and a wavelength
of 32 pc (see also Gomez de Castro \& Pudritz 1992).
To summarize, periodic structures with wavelengths in the
range 3 -- 30 pc are often found in molecular clouds.

The solutions presented in this paper represent stationary
waves in the molecular cloud fluid.  Fortunately, a characteristic
observable signature of these stationary waves is the relation given
by the continuity equation: $\rho v = A$.  If we observe a candidate
wave train in a molecular cloud, then the observed line-center velocity
(along the propagation direction of the wave) should vary
inversely with the density.  Notice also that the linewidth is known
to vary with density (see the discussion following equation [2.4]).
Since the linewidths decrease with increasing density, the observed
linewidths, when measured along the direction of a wavetrain,
should also be anticorrelated with the density.  Keep in mind,
however, that this linewidth variation is a signature of the
equation of state, whereas the aforementioned line-center velocity
variation is a signature of stationary waves.

One example of stationary wave behavior in an observed molecular
cloud may be provided by the Lynds 204 complex (McCutcheon et al. 1986).
This cloud is highly elongated and shows periodic structure
in column density with a wavelength of $\sim 1^\circ$ ($\sim$3 pc
for an assumed distance to the cloud of 160 pc).  In addition,
these authors find that the line of sight velocities (of CO line
observations) along the filament are very well correlated with
the density enhancements in the sense that the most massive
parts of the filament have the smallest velocity displacement
from the mean $v_0$ (see especially their Figures 4 and 5).
Finally, the authors derive the relation $M_j (v_j - v_0)$ =
{\it constant}, where $M_j$ and $v_j$ are the mass and velocity
of the $j$th segment of the filament.  This relation looks
suspiciously like the solution to the one dimensional continuity
equation for a stationary wave (see equation [2.17]).

We have also found that nonlinear stationary waves select out
particular lengthscales for their wavelengths (see, e.g.,
Figures 4, 5, and 8).  The observed wavelength should thus
be a calculable function of the various
physical parameters involved (wave amplitude, sound speed, magnetic
field strength, etc.).  Although several parameters are necessary to
completely specify the preferred wavelength (or range of wavelengths),
all of the parameters are physical quantities and can be determined,
in principle, independently of the the observations of the wavelengths.

Another observable quantity is the relative widths of the waves.
The candidate wavetrains in molecular cloud regions appear as
chains of ``blobs'' which are (more or less) lined up and regularly
spaced.  The aspect ratio of these blobs (at a chosen density
contour) is thus a well-defined observable quantity (this aspect
ratio has a value of $3 \pm 1$ for the sample of Schneider
\& Elmegreen 1979). We are currently studying stationary waves
on filaments and can obtain a prediction for the
relative widths of the wavetrains (Gehman et al. 1993) as a
function of the physical parameters of the problem.

In addition to forming substructure within molecular clouds,
wave motions might be partly responsible for the formation of
molecular clouds themselves.  Consider, for example, a large
scale nonlinear wavetrain (with a wavelength $\lambda$ of
several pc) traveling through low density ($n \sim$ 1)
atomic gas.  Atomic regions with sufficiently large
column densities can become self-shielding and
transform into regions containing molecular gas.
After the wave has passed by, a wake of molecular material
can remain. This process would thus leave behind a permanent
record of the passage of the wave. We note that somewhat
similar behavior forms clouds in the Earth's atmosphere,
where waves set up regions of both low and high density;
in the regions of high density, moisture condenses to form clouds
and the so-called herring-bone cloud structure is produced
(see, e.g., the introduction of Infeld \& Rowlands 1990).
We note that molecular cloud formation probably involves many
different physical processes (see, e.g., the review of Elmegreen
1987).  However, the large scale waves discussed here may play
an important role and should be studied further in this context.

\medskip
\centerline{\it 8.3 Discussion and Directions for Future Work}
\medskip

This paper (see also Paper I and AFW)  provides a first step
toward an understanding of wave motions and structure formation
in molecular clouds. However, much more work remains to be done.
Perhaps the key point that one should keep in mind is that complex
fluid systems, such as molecular clouds, can wiggle around in many
different ways.  In this paper, we have concentrated on the study of
stationary waves (travelling waves of permanent form) in one spatial
dimension.  Within this class of waves, a great variety of solutions
exists.  However, an even greater variety of wave motions can arise
from the time dependent problem. These more general types of wave
solutions should be studied.

Another related topic is the actual generation of the wave
motions.  As discussed in Paper I, we expect that
self-gravitating clouds will tend to collapse and excite a
wide spectrum of wave motions (see, e.g., Arons \& Max 1975).
The self-gravity of the cloud can provide a more than
adequate energy source for the waves.  Any wave motions
which get excited will generally either grow or disperse.
However, the waves which live the longest will be the waves
of permanent form -- the stationary waves studied in this
paper.   One classic example of this type of scenario was
studied by Zabusky \& Kruskal (1965).  They considered the
time evolution of the Korteweg-deVries equation (a model
equation which describes, among other things, surface waves
on water; see Korteweg \& de Vries 1895) and showed that
initially simple cosine waves develop into a train of highly
nonlinear pulses of permanent form; these pulses are the
classic example of true (stable) solitons.  Analogous calculations
should be performed for the self-gravitating fluid systems of
this paper and for more detailed models which more closely
resemble real molecular clouds.

Wave stability provides another avenue for future research.
Even though the waves considered here are stationary and thus
possess a balance between gravitational dispersion and nonlinear
steepening, they may be subject to instabilities (see, e.g.,
Pego \& Weinstein 1992). In the likely event that the waves are
unstable, we must find the wave configurations with the longest
lifetimes; these waves will be the ones which provide, in part,
the observed structure of molecular clouds.

\vskip 0.6truein
\centerline{Acknowledgements}
\medskip

We would like to thank Phil Myers, Ira Rothstein, Scott Tremaine,
Michael Weinstein, and Jennifer Wiseman for useful discussions.
We also thank an anonymous referee for many useful comments.
This work was supported by NASA Grant No. NAGW--2802,
the NSF Young Investigator Program, and funds from the
Physics Department at the University of Michigan.
MF is supported by a Compton GRO Fellowship from NASA.

\bigskip
\bigskip
\centerline{\bf APPENDIX A: SCALING TRANSFORMATIONS}
\medskip

In this Appendix, we consider a class of transformations which
leave the fundamental equation of motion of this theory
invariant.  We begin with the equation itself:
$$\rho \rho_{\xi \xi} \Bigl[ \rho^2 \dpdrho - A^2 \Bigr] +
\rho_\xi \rho_\xi \Bigl[ 3 A^2 - \rho^2 \dpdrho
+ \rho^3 {\partial^2 p \over \partial \rho^2} \Bigr]
+ \rho^4 q (\rho)  = 0 \, , \eqno(\scale1)$$
which is written with an arbitrary charge density.
We now consider transformations of the density $\rho$
and the variable $\xi$ of the form
$$\rho \to \Lambda \rho \, , \eqno(\scale2)$$
$$\xi \to \gamma \xi  \, . \eqno(\scale3)$$
The equation of motion [A1] remains invariant under this
transformation provided that
$$\gamma = \Lambda^{-1/2} \, , \eqno(\scale4)$$
and the constituents of the problem scale according to
$$ A \to \Lambda A \, , \eqno(\scale5)$$
$$ p \to \Lambda p \, , \eqno(\scale6)$$
$$ q \to \Lambda q \, . \eqno(\scale7)$$
The Mach number $A$ and the parameters in the pressure $p$ can
generally be  rescaled as required.  In order for the scaling
of the charge density (equation [A7]) to be satisfied,
the parameters of the charge density must also be
rescaled appropriately, as we discuss below.

For Jeans theory, $q = \rho - \rho_0$, and the rescaling of
equation [A7] can be met if $\rho_0 \to \Lambda \rho_0$.
In practice, we choose $\Lambda$ to make $\rho_0 = 1$ and
then rescale the remaining parameters as described above.

Both the Yukawa theory and the pseudo-two-dimensional theory
of \S 7 can also be rescaled.  We require
$$m^2 \to \Lambda m^2 \, , \eqno(\scale8)$$
for the former and
$$\ypro^2 \to \Lambda \ypro^2 \, , \eqno(\scale9)$$
for the latter.  These scalings are not unexpected since
both $m$ and $\ypro$ represent inverse lengthscales
(very roughly they are the wavenumbers in the direction
perpendicular to the propagation direction of the wave)
and since lengthscales transform as in equations [A2]
and [A3].

\bigskip
\bigskip
\centerline{\bf APPENDIX B: SOLITARY WAVE SOLUTIONS IN JEANS THEORY}
\medskip

In this Appendix, we consider a class of solitary wave solutions
in Jeans theory (see \S 4 in the text). For these solutions, the
singularity is removed by choosing $\beta$ such that
$f(\rho_C)=0$.  We first expand $f$ (see equation [2.20])
about $\rho_C$ to obtain
$$f =  {(\rho-\rho_C)^2\over 2} {(\rho_0 - \rho_C) \over
\rho_C^3} \left\{{\partial\over\partial\rho}\left(\rho^2{\partial p\over
\partial\rho}\right)\right\}_C +  {\cal O}
\bigl[ (\rho-\rho_C)^3 \bigr]  , \eqno({\rm B}1)$$
where the $C$ subscript represents evaluation of the term in
brackets at $\rho_C$.  We now expand equation [B1] about
the point $\rho_C$ and obtain
$$
{1\over 2}\rho^2_\xi = {\rho_C^3\over 2} \, (\rho_0 - \rho_C)
\left[\left\{{\partial\over\partial\rho}\left(\rho^2{\partial p\over
\partial\rho}\right)\right\}_C\right]^{-1} +  {\cal O}
\bigl[ (\rho-\rho_C) \bigr]  \, . \eqno({\rm B}2)$$
For the particular equation of state used here (see equation
[2.4b]), the expansion reduces to a simpler form
$$
{1\over 2}\rho^2_\xi = {\rho_C^3\over 2} \, (\rho_0 - \rho_C)
\left[ 2 \rho_C + \kappa \right]^{-1} +  {\cal O}
\bigl[ (\rho-\rho_C) \bigr]  \, . \eqno({\rm B}3)
$$
Since the term in square brackets is nonzero,
and since $f \ge 0$ near $\rho_C$, the quantity
$\rho_\xi^2$ must be positive and nonsingular.

For physically valid solutions, the density must lie in the
range between $\rho_1 = 0$ and $\rho_2$ (where the peak
density obeys the ordering $\rho_2 > \rho_0 > \rho_C$).
The first integral of the equation of motion (see equation
[2.19]) may be expanded about $\rho = \rho_1 = 0$ to obtain
$$\rho^2_\xi = {\rho_0 \over  A^2} \rho^4 + {\cal O}
\bigl( \rho^5 \bigr)  . \eqno({\rm B}4)$$
Thus, in the limit $\rho \to \rho_1=0$, the wave profile
has the form
$$\rho \approx \left[{A^2\over\rho_0}\right]^{1/2}\;
{1\over |\xi|} \;.\eqno({\rm B}5)$$
It is evident from equation [B5] that this class of solutions
represents solitary waves (see the discussion in \S 2.6).
The density profile $\rho(\xi)$ for this case is shown in Figure 6.
The dashed lines correspond to the sonic points. This solution
corresponds to an unshocked flow with both supersonic and
subsonic regimes and with densities both above and below the
``background'' density $\rho_0$.

One complication of this solution is that the boundary conditions
in the limits $\xi \to \pm \infty$ are unphysical because the velocity
$v\sim\rho^{-1}\sim\xi$ diverges.  Furthermore, the total charge
$Q$ also diverges, as shown by substituting the limiting form of
$f(\rho\to 0)\to A^2\rho_0/2\rho^2$ into equation [3.8].
In order for this solution to be physical, appropriate boundary
conditions must be applied at some finite distance.
Suppose we interpret the initial unperturbed fluid to be at
uniform density $\rho_0$ and to have zero initial charge.
The nonlinear structures that form out of this configuration
must also have zero total charge.  Thus, one ``natural''
choice of boundary conditions is to assume a symmetric density
profile that also has zero total charge, i.e., we cut off the
density profile at a finite lengthscale such that the total
charge inside vanishes.  As shown by equation [3.8], the
$Q=0$ solitary wave is bounded by the sonic points
(since $f(\rho_C)=0$).

It is worth noting that charge densities $q(\rho)$ can exist for
which these solutions (i.e., solutions in which the singularity is
removed in this manner) produce wave profiles which are bounded
by a lower density $\rho_1 \ne 0$, thereby allowing physical
boundary conditions at spatial infinity.

\bigskip
\bigskip
\centerline{\bf REFERENCES}
\medskip

\par\pp
Adams, F. C. 1992, {\sl ApJ}, {\bf 387}, 572

\par\pp
Adams, F. C., \& Fatuzzo, M. 1993, {\sl ApJ}, {\bf 403}, 142 (Paper I)

\par\pp
Adams, F. C., Fatuzzo, M., \& Watkins, R. 1993, submitted to
{\sl Phys. Lett.}, {\bf A}, (AFW)

\par\pp
Arons, J. \& Max, C. 1975, {\sl ApJ}, {\bf 196}, L77

\par\pp
Binney, J., \& Tremaine, S. 1987, Galactic Dynamics
(Princeton: Princeton University Press)

\par\pp
Blitz, L. 1993, in Protostars and Planets III,
ed. E. Levy and M. S. Mathews (Tucson: University of
Arizona Press), in press

\par\pp
Coleman, S. 1985, Classical Lumps and their Quantum
Descendants, in Aspects of Symmetry (Cambridge:
Cambridge University Press)

\par\pp
Dame, T. M., Elmegreen, B. G., Cohen, R. S., \& Thaddeus, P. 1986,
{\sl ApJ}, {\bf 305}, 892

\par\pp
de Geus, E. J., Bronfman, L., \& Thaddeus, P. 1990, {\sl A \& A},
{\bf 231}, 137

\par\pp
Dewar, R. L. 1970, {\sl Phys. Fluids}, {\bf 13}, 2710

\par\pp
Drazin, P. G., \& Johnson, R. S. 1989, Solitons: An Introduction
(Cambridge: Cambridge University Press)

\par\pp
Elmegreen, B. G. 1987, in Interstellar Processes, ed.
D. J. Hollenbach and H. A. Thronson (Dordrecht: Reidel), p. 259

\par\pp
Elmegreen, B. G. 1990, {\sl ApJ}, {\bf 361}, L77

\par\pp
Gehman, C., Adams, F. C., Fatuzzo, M., \& Watkins, R. 1993, in preparation

\par\pp
Goldsmith, P. F., \& Arquilla, R. 1985, in Protostars and
Planets II, eds. D. C. Black and M. S. Mathews (Tucson:
Univ. Arizona Press), p. 137

\par\pp
Gomez de Castro, A. I., \& Pudritz, R. E. 1992,
{\sl ApJ}, {\bf 395}, 501

\par\pp
Goodman, A. A., Crutcher R. M., Heiles, C., Myers, P. C., \&
Troland, T. H. 1989, {\sl ApJ}, {\bf 338}, 161

\par\pp
G\"otz, G. 1988, {\sl Class. Quantum Grav.}, {\bf 5}, 743

\par\pp
Houlahan, P., \& Scalo, J. 1992, {\sl ApJ}, {\bf 393}, 172

\par\pp
Infeld, E., \& Rowlands, G. 1990,
Nonlinear Waves, Solitons, and Chaos
(Cambridge: Cambridge University Press)

\par\pp
Jeans, J. H. 1928, {\sl Astronomy and Cosmogony} (Cambridge:
Cambridge Univ. Press)

\par\pp
Korteweg, D. J., \& de Vries, G. 1895, {\sl Philos. Mag.},
{\bf 39}, 422

\par\pp
Langer, W. D. 1978, {\sl ApJ}, {\bf 225}, 95

\par\pp
Larson, R. B. 1981, {\sl MNRAS}, {\bf 194}, 809

\par\pp
Liang, E. P. T. 1979, {\sl ApJ}, {\bf 230}, 325

\par\pp
Lizano, S., \& Shu, F. H. 1989, {\sl ApJ}, {\bf 342}, 834

\par\pp
McBreen, B., Fazio, G. G., Stier, M., \& Wright, E. L. 1979,
{\sl ApJ}, {\bf 232}, L183

\par\pp
McCutcheon, W. H., Vrba, F. J., Dickman, R. L., \& Clemens, D. P.
1986, {\sl ApJ}, {\bf 309}, 619

\par\pp
Mouschovias, T. Ch. 1976, {\sl ApJ}, {\bf 206}, 753

\par\pp
Mouschovias, T. Ch. 1978, in Protostars and Planets,
ed. T. Gehrels (Tucson: University of Arizona Press),
p. 209

\par\pp
Myers, P. C. 1983, {\sl ApJ}, {\bf 270}, 105

\par\pp
Myers, P. C. 1987, in {Interstellar Processes},
ed. D. Hollenbach and H. Thronson,
(Dordrecht: Reidel), p. 71

\par\pp
Myers, P. C., \& Goodman, A. A. 1988, {\sl ApJ}, {\bf 329}, 392

\par\pp
Myers, P. C. 1991, in {Fragmentation of Molecular Clouds
and Star Formation} (IAU Symposium 147), eds. E. Falgarone and
G. Duvert (Dordrecht: Kluwer), in press

\par\pp
Myers, P. C., \& Fuller, G. A. 1992, {\sl ApJ}, {\bf 396}, 631

\par\pp
Nakano, T. 1985, {\sl Pub. A.S.J.}, {\bf 37}, 69

\par\pp
Pego, R. L., \& Weinstein, M. I. 1992, {\sl Phys. Lett. A},
{\bf 162}, 263

\par\pp
Pudritz, R. E. 1990, {\sl ApJ}, {\bf 350}, 195

\par\pp
Rajaraman, R. 1987, Solitons and Instantons: An Introduction
to Solitons and Instantons in Quantum Field Theory (Amsterdam:
North Holland)

\par\pp
Ray, D. 1983, {\sl J. Math. Phys.}, {\bf 24}, 1011

\par\pp
Riemann, B. 1858, G\"ottingen Abhandlungen, Vol. viii, p. 43

\par\pp
Russell, J. S. 1844, 14th Meeting of the British Association
Report (York), p. 311

\par\pp
Schneider, S., \& Elmegreen, B. G. 1979, {\sl ApJ}, {\bf 41}, 87

\par\pp
Shu, F. H. 1983, {\sl ApJ}, {\bf 273}, 202

\par\pp
Shu, F. H. 1992, Gas Dynamics (Mill Valley: University Science Books)

\par\pp
Shu, F. H., Adams, F. C., \& Lizano, S. 1987, {\sl ARA\&A}, {\bf 25}, 23

\par\pp
Shuter, W. L., Dickman, R. L., \& Klatt, C. 1987,
{\sl ApJ}, {\bf 322}, L103

\par\pp
Whitham, G. B. 1974, Linear and Nonlinear Waves
(New York: Wiley)

\par\pp
Wiseman, J. J., \& Adams, F. C. 1993, in preparation

\par\pp
Wood, D.O.S., Myers, P. C., \& Daugherty, D. A. 1993,
to be submitted to {\sl ApJ Suppl.}

\par\pp
Zabusky, N. J., \& Kruskal, M. D. 1965, {\sl Phys. Rev. Lett.},
{\bf 15}, 240

\vskip 0.5truein
\bigskip
\centerline{\bf FIGURE CAPTIONS}
\medskip

\noindent
Figure 1. Schematic of the function $f(\rho)$ showing possible
behavior of a charge density theory in which $q(\rho)$ has at
least two zeroes.  The solid curve shows the usual case in which
a second ordinary zero of $f$ exists.  The long dashed curve shows
the case in which $f(\rho)$ asymptotically approaches a constant
and hence no soliton solutions are allowed. For the remaining
case shown by the short dashed curve, depression solitons are allowed.

\medskip
\noindent
Figure 2. Schematic of the function $f(\rho)$ for Jeans theory.

\medskip
\noindent
Figure 3. Wave profile $\rho(\xi)$ for nonlinear
waves in Jeans theory.  This example has $A = 1$ and
$\kappa = 10$.

\medskip
\noindent
Figure 4. Allowed range of wavelengths for nonlinear
waves in the Jeans theory.  The upper curve shows the
(nonlinear) Jeans wavelength $\lambda_J$ plotted as a
function of amplitude $\delta \rho/\rho$ of the wave.
The lower curve shows the ``shock limit'', i.e., waves
with wavelengths smaller than this critical value will
shock the dissipate.  The wavelengths are expressed
in units of the linear Jeans length for the pressure
given by equation [2.4b] with $\kappa$ = 10.
Nonlinear waves in this theory must have wavelengths
which lie between the two curves.

\medskip
\noindent
Figure 5. Allowed range of wave profile widths $\lambda_p$ for
nonlinear waves in the Jeans theory.  The upper curve shows the
width for the longest possible wavelength (the Jeans wavelength)
plotted as a function of amplitude $\delta \rho/\rho$ of the wave.
The lower curve shows the width for the ``shock limit''.
All quantities are the same as in Figure 4.
The width of the wave profiles for nonlinear waves
in this theory are thus confined to a rather narrow range.

\medskip
\noindent
Figure 6. Solitary wave solution for Jeans theory for $A$ = 3
and $\kappa$ = 10.  Vertical dashed lines show the location of
the sonic points (see \S 4 and Appendix B).

\medskip
\noindent
Figure 7. Wave profile $\rho(\xi)$ for nonlinear waves in
Jeans theory including the effects of magnetic fields as a pressure
term.  This example has $A$ = 1, $\kappa$ = 10, and $\alpha$ = 3;
the constant of integration $\beta$ has been adjusted to make
the amplitude comparable to the field-free case shown in Figure 3.
Notice the enhanced width of the waves relative to those of
Figure 3.

\medskip
\noindent
Figure 8.  Allowed range of wavelengths for nonlinear waves in the
Jeans theory with magnetic fields included in the flux freezing
approximation.  The upper curve shows the (nonlinear) Jeans
wavelength $\lambda_J$ plotted as a function of amplitude
$\delta \rho/\rho$ of the wave.
The lower curve shows the ``shock limit'', i.e., waves
with wavelengths smaller than this critical value will
shock and dissipate.  Here, $\kappa$ = 10, $\alpha$ = 3,
and the wavelengths are normalized as in Figure 4.
Nonlinear waves in this theory
must have wavelengths which lie between the two curves.

\medskip
\noindent
Figure 9. Wave profiles $\rho(\xi)$ for a solitary wave
(upper curve) and a ``depression soliton'' (lower curve)
in Yukawa theory.  Both examples have $A$ = 1 and $\kappa$ = 10.
Notice that the depression soliton has a much lower density
constrast than the ordinary solitary wave.

\medskip
\noindent
Figure 10. Allowed regions of the $A-m$ plane for various types
of stationary waves (for Yukawa theory and $\kappa$ = 10).
Solitary wave solutions are allowed in both regions I and II.
Depression solitons are limited to region II.  Ordinary
nonlinear stationary waves are allowed in region III.
The vertical dashed curve on the right side of the figure
corresponds to the sonic point for this equation of state;
all allowed Mach numbers $A$ must be less than this value.

\bye